\documentclass{ws-ijmpd}
\usepackage[super,compress]{cite}
\usepackage{graphicx}
\usepackage{xcolor}
\usepackage{hyperref}

\begin{document}

\markboth{A. Cuevas, J.Chagoya, C.Ortiz .}
{Boundary terms in cosmology}

\catchline{}{}{}{}{}
%
%%%%%%%%%%%%%%%%%%%%%%%%%%%%%%%%%%%%%%%%%%%%%%%%%%%%%%%%%%%%%%%%%%%%

\title{Boundary terms in cosmology}

\author{A. Cuevas, Javier Chagoya and C. Ortiz.}

\address{Unidad Académica de Física, Universidad Autónoma de Zacatecas, Calzada  Solidaridad esquina con Paseo a la Bufa S/N\\
Zacatecas, Zacatecas C.P. 98060, México\\
angel.cuevas@fisica.uaz.edu.mx}
\maketitle

\begin{history}
\received{Day Month Year}
\revised{Day Month Year}
\end{history}

\begin{abstract}
% Due the Cauchy problem in general relativity and the Einstein constrain equations, the Einstein field equations are well-posed. 
%In the derivation of the Einstein field equations from Hamilton's principle, a boundary term is required to make the variational problem well-posed. This boundary term appears as a counterterm, for instance the Gibbons-Hawking-York term, that cancels out variations that do not necessarily vanish at the boundary of the spacetime manifold. In this work we explore an alternative way of dealing with the boundary term. In a cosmological setting, we use the Lagrange multiplier method to force the vanishing of the boundary term at all times. As a result we predict the existence of a fluid that decays as the sixth power of the scale factor. For different reasons, this type of fluid and its consequences for the early universe have been considered before in the literature under the name of \textit{kination}.
%is usually dropped out by the Stoke's theorem, and usually one need to add a counterterm to be the action well-posed. \\ In this paper we use the Lagrange multiplier method to add this boundary term as a \textit{constrain equation} or \textit{force of constrain} into the Einstein field equations and drop the boundary term as an auxiliary condition. As a result we obtain a prediction over the Friedmann equation, a fluid that decays like the inverse of the scale factor to the six power. This type of fluid (or scalar field) is know as \textit{kination} wich could be a era in the past of the \textit{whole Universe}.
In the derivation of the Einstein field equations via Hamilton’s principle, the inclusion of a boundary term is essential to render the variational problem well-posed, as it addresses variations that do not vanish at the boundary of the spacetime manifold. Typically, this term is chosen as the Gibbons-Hawking-York boundary term. %., which acts as a counter-term to cancel out the undesirable boundary contributions. 
In this work, we propose an alternative treatment of the boundary term within a cosmological framework by employing the Lagrange multiplier method. This approach enforces the vanishing of the boundary term throughout the evolution of the Universe, leading to the prediction of a  fluid component that decays as the sixth power of the scale factor. This type of fluid has been studied in the context of the early universe under the name of \textit{stiff matter}, and it can be related to a scalar field known as \textit{kination}.

\end{abstract}


\ccode{PACS numbers:}

%\tableofcontents

\section{Introduction}	
The variational principle  is a fundamental tool for deriving the equations of motion for physical systems. In the context of mechanics, it is well known that integration by parts allows one to separate the variational expression into a \textit{bulk} term and a \textit{boundary} term. For many physical theories, boundary terms play a crucial role; examples include Dirichlet and Neumann boundary conditions, as well as the concepts of Noether’s current and Noether’s charge, all of which emerge from boundary contributions \cite{Krishnan:2016mcj, Izumi:2023rwh, DeHaro:2021gdv}.

David Hilbert adopted the \textit{Hamilton’s principle}, or \textit{principle of least action}, to derive the Einstein field equations (EFE) from a variational framework \cite{Hilbert1915}. This principle of least action identifies a path that nature ‘selects’ for physical processes \cite{lanczos1986variational}. Hamilton’s principle, fundamentally an extremal principle, applies strictly to monogenic forces (those derivable from a potential) which include only scleronomic holonomic constraints. However, extensions to Hamilton’s principle exist to handle systems with semiholonomic constraints, incorporating generalized forces to produce well-defined equations of motion in classical mechanics \cite{goldstein2002classical}.
In the variational process, the boundary terms are typically omitted through the application of \textit{Stokes’ theorem} \cite{gron2007einstein, Carroll:2004st}, simplifying the derivation of equations that govern {dynamical} systems.

In General Relativity (GR), achieving a well-posed action principle often {requires} adding appropriate  counterterms to address boundary contributions effectively. One such term is the Gibbons-Hawking-York (GHY) term; however, there are other counterterms that play the same role. This ambiguity underscores the need for a systematic analysis and selection of boundary terms in gravitational theories \cite{Parattu:2016trq, Chakraborty:2016yna}. 

We propose an alternative approach that involves using Lagrange multipliers to %dynamically 
enforce the vanishing of these boundary contributions. Incorporating the Lagrange multiplier method allows to systematically determine the {constraints} due to boundary terms. This leads to modified equations of motion that inherently account for boundary effects. This method not only generalizes the treatment of boundary terms but it may also uncover new dynamical properties, as we will show in the following by considering a cosmological scenario.%such as the emergence of unique fluid components, as seen in cosmological models where these constraints influence the evolution of the universe.

%The Variational Principle (VP) is one of the most useful techniques to obtain the equations of motion of a physical system. In the variational principles of the mechanics, it's is well know that by integration by parts, one can dive the problem into a \textit{bulk} term and a \textit{boundary} term (BT). 
 %For many theories this boundary terms may a important role. Dirichlet boundary conditions, Von Nwemann boundary conditions, Noether's current and Noether's charge, etc; all of these mathematical and physical terms come from the boundary terms \cite{Krishnan:2016mcj, Izumi:2023rwh ,DeHaro:2021gdv}.\\ David Hilbert, was adoptep the \textit{Hamilton's principle} or \textit{principle of least action} to derive the Einstein field equations (EFE) from a variational principle. The principle of least action involves a path that nature tends to choose in physical process \cite{lanczos1986variational}. 
  %In the process of variation, the boundary terms are usually left out by Stoke's theorem \cite{gron2007einstein, Carroll:2004st}. 
  
  %In GR it can usually be added a counter term like the GHY counter term to obtain a well posed action \cite{Parattu:2016trq}\cite{Chakraborty:2016yna}. \\Hamilton's principle is a minimum principle, it's only valid for monogenic forces that are \textit{monogenic in nature}. There are extension to hamilton principle to handle with non-holonomic constrains \cite{goldstein2002classical}. These equations of motion are provided with a generalized force in classical mechanics. 
Building upon this approach, we now consider a cosmological setting similar to the cosmological standard model used in GR, where the matter and energy content of the Universe is modeled with a perfect fluid characterized by its local energy density $\rho$ (in units $c=1$), pressure $P$ and by an equation of state $P = \omega \rho$. The parameter $\omega$ takes specific values for different types of fluids: $\omega = -1 $ corresponds to a \textit{cosmological constant} which is introduced as a model of \textit{dark energy}, $\omega = 0$  for \textit{dust} (non-relativistic matter), and $\omega = 1/3$  for \textit{radiation}. However, $\omega$ can also take other values, as in the cases of \textit{quintessence }and \textit{phantom} fluids. For quintessence, $\omega$ typically lies in the range $-1<\omega\leq-1/3$  representing a dynamical scalar field whose effective pressure is negative and weaker than that of the cosmological constant. On the other hand, \textit{phantom} fluids have $\omega < -1$ {and correspond to another} form of dark energy with even more negative pressure, which can lead to a \textit{Big Rip} scenario.
{Another fluid that is central in the present work  is known as \textit{stiff matter}}, $\omega = 1$, a type of fluid theorized by Ya. B. Zel'dovich where he describes the most rigid equation of state compatible with the requirements of the theory of relativity\cite{Zeldovich:1961sbr}. An alternative representation of this fluid, in the post-inflationary era, is described by a fast-rolling scalar field, known as \textit{kination}, firstly theorized by Michael Joyce. \cite{Joyce:1996cp} This scalar field was considered to explain the production of baryons at electroweak phase transition and dominates at an era of the Universe when the scalar field only had kinetic energy.

This paper is organized as follows: In Sec.~\ref{sec:2}, we present a brief notion of boundary terms and its relation with well-posed problems. Also, we describe the variation of the GR action and the boundary terms of the theory. Finally, we evaluate the geometrical \textit{cosmological boundary term} and we describe how this term is set to vanish. 
In Sec.~\ref{sec:3}, we use the Lagrange multiplier method in a cosmological scenario and we derive the evolution equations of the model, finding novel predictions.
Finally, Sec.~\ref{sec:4}, is devoted to discussion and conclusions of this paper.
%%%%%%%%%%%%%%%%%%%%%%%%%%%%%%%%
\section{Boundary terms}\label{sec:2}
%%%%%%%%%%%%%%%%%%%%%%%%%%%%%%%%
In the framework of the variational principle, boundary terms play a crucial role in ensuring that the derived equations of motion are consistent and well-posed. From a more general perspective, well-posedness in partial differential equations (PDEs) means that all problems involving differential equations require initial conditions to ensure a unique solution \cite{hadamard1902problemes}. In the calculus of variations, the elementary theory of PDEs necessitates the inclusion of boundary conditions to define a well-posed problem. These boundary conditions, which may be determined by physical circumstances, can be classified into two types:
\begin{enumerate}
\item Imposed boundary conditions. These are conditions set externally by the experimenter or theorist.
\item Natural boundary conditions.  These are conditions that arise inherently from the variational problem itself.
\end{enumerate}
By considering both types of boundary conditions, the problem’s solution becomes unique (see{, e.g.,} ``The calculus of variations and boundary conditions: The problem of the elastic bar" \cite{lanczos1986variational}). In the context of GR, boundary terms are essential for ensuring a well-posed variational problem, as spacetime boundaries can have significant contributions to the dynamics. Typically, boundary terms are introduced to cancel out variations that do not vanish at the boundary, with the GHY\cite{York:1986lje,Gibbons:1976ue} term being a well-known example.  The EFE are a well-posed Cauchy problem due to the constraint equations in GR \cite{chrusciel2004einstein, Karp:2009tq, Isenberg:2013iva}.  %In addition to a well-posed variational principle, one also needs a well-defined system of differential equations. For instance, 
In cosmology, %the FLRW metric is the solution to the EFE that describes the Universe with high accuracy at large scales. 
in order to determine the evolution of the Universe, it is required to impose appropriate   \textit{initial conditions} or \textit{boundary conditions}.

%The universe, as a \textit{whole local universe}, requires the inclusion of \textit{initial conditions} or \textit{boundary conditions}; otherwise, it would not constitute a well-posed problem related to PDEs.

%The Lagrange multiplier method is a tecnique that reduce degrees of freedom that evolves certain \textit{constrain equations} of a mathematical problem. In physics the Lagrange  multipliers are related to the \textit{reaction forces} of a mechanical problem. So a most general comprension of a mechanical problem need this form of treathment.\\
%Also is know that this method applicated in VP is still avaliable for a \textbf{non-holonomic}, \textbf{semi-holonomic}, and  \textbf{holonomic} constrains\cite{lanczos1986variational, courant89 }.\\ In the equations of motion the generalized force due to constrains is know as \textit{constraint force} \cite{josé1998classical}.\\ \\

%%%%%%%%%%%%%%%%%%%%%%%%%%%%%%%%%%%%%%%%%%%
\subsection{{Boundary terms for the action of general relativity}}
%%%%%%%%%%%%%%%%%%%%%%%%%%%%%%%%%%%%%%%%%%%
In general relativity, the derivation of the equations of motion from the variational principle naturally gives rise to boundary terms during the variation of the action. These terms, which result from integration by parts, are often eliminated using Stokes’ theorem. However, a careful treatment of these boundary terms can provide valuable insights into the underlying dynamics of the system.
%In deriving the equations of motion from the variational principle in GR, boundary terms naturally appear during the variation of the action. These terms, arising from integration by parts, are typically discarded using Stokes' theorem. However, careful treatment of boundary terms can offer deeper insights into the system's dynamics.
 Let us review the derivation of the EFE, highlighting the role of the boundary terms. We begin with the  action,
%The following derivation begins with the GR action, highlighting the role of these boundary terms in the derivation of EFE.
\begin{equation}
           S=  S_{EH} + S_M =  \int d^4x \sqrt{-g} \left(\frac{R}{2\kappa} + \mathcal{L_M}\right)  ,
       \end{equation} 
where $\kappa = 8 \pi G $ is the Einstein's gravitational constant and $\mathcal {L_M}$ is the matter Lagrangian density.  The equations of motion are derived through the principle of least action, {leading to} 
    \begin{equation}
          0=  \frac{1}{2\kappa} \int d^4x \sqrt{-g}\left[  R_{\alpha \beta} - \frac{1}{2} R  g_{\alpha \beta}  - {\kappa} T_{\alpha \beta} \right]\delta g^{\alpha \beta}  +\int  \frac{d^4x}{2\kappa} \partial_\sigma (\sqrt{-g}B^\sigma) + \int d^4x\partial_\sigma G^\sigma \label{GRvar}.
        \end{equation}
         Assuming that in general the matter Lagrangian depends on the metric and its first derivatives \cite{gron2007einstein},
         \begin{equation}
        T_{\alpha\beta} = \frac{-2}{\sqrt{-g}}\left(\frac{\partial (\sqrt{-g} \mathcal{L_M})}{\partial g^{\alpha \beta}}  - \partial_\sigma  \frac{\partial (\sqrt{-g} \mathcal{L_M})}{\partial (\partial_\sigma g^{\alpha \beta})} \right),
     \end{equation} 
     is defined as the \textit{energy-momentum tensor.} \\The vector density $G^\sigma$ and the vector $B^\sigma$ are given by:
\begin{align} G^\sigma & = 
    \frac{\partial \left(\sqrt{-g} \mathcal{L_M}\right)}{\partial (\partial_\sigma g^{\alpha \beta})} \delta g^{\alpha \beta}.
\\
         B^\sigma & = g^{\alpha \beta} \delta \Gamma^\sigma_{\beta \alpha} - g^{\alpha \sigma}\delta \Gamma^\beta_{\beta \alpha}.
     \end{align}
     The last two integrals in Eq. \eqref{GRvar} are the \textit{divergence terms} or \textit{boundary terms} of the theory. % These are usually dropped out by Stoke's theorem \cite{gron2007einstein, Carroll:2004st}. 
For a perfect fluid there is no known  matter Lagrangian term that depends on the metric derivatives, therefore $G^\sigma=0$. For the geometrical boundary term, it is usual to add the GHY counter-term to cancel out $B^\sigma$. 
     By requiring that the variation of the total action vanishes for arbitrary variations of the metric, we obtain the \textit{Einstein field equations}
    \begin{equation} 
     R_{\alpha \beta} - \frac{1}{2} R g_{\alpha \beta} = \kappa T_{\alpha \beta},
         \end{equation}
where $R_{\alpha\beta}$ is the \textit{Ricci tensor} and $R$ is the \textit{Ricci scalar}.
%%%%%%%%%%%%%%%%%%%%%%%%%%%%%%%%%%%%%%%%%%%%%%%%%
\subsection{Cosmological boundary term} 
%%%%%%%%%%%%%%%%%%%%%%%%%%%%%%%%%%%%%%%%%%%%%%%%%%%
From the \textit{Copernican principle}, if the Universe appears isotropic about our position (in the Earth), it would be  isotropic for other observers in every other places in the Universe. Therefore,  it must be homogeneous everywhere \cite{Peacock:1999ye}.
Following the Copernican principle, since we observe an isotropic Universe at large scales, we can assume that the Universe is homogeneous at those scales. %say that at large scales, the Universe looks \textit{homogeneous} and \textit{isotropic}.
The Friedmann-Lemaitre-Robertson-Walker (FLRW) metric (a spatial maximally symmetric spacetime) is a solution to the EFE which describes the aforementioned characteristics. The FLRW line element is
\begin{equation}
    ds^2 = -N(t)^2dt^2 + a(t)^2\left(\frac{dr^2}{1-kr^2} + r^2 d\theta^2 + r^2 \sin{\theta}^2 d\phi ^2\right),
\end{equation}
with $N(t)$ as the \textit{lapse function}, $a(t)$ the \textit{scale factor}, and $k$ representing the spatial curvature.
In this spacetime, choosing a boundary fixed at constant $r$, fixing $\theta = \frac{\pi}{2}$ without loss of generality, and assuming a spatially flat FLRW universe, \textit{i.e.}, $ k= 0$, the only non-vanishing component of the geometrical boundary term $B^\sigma$ is 
\begin{equation}
    B^0 = \frac{3}{2} \left(\frac{1}{a^2}\right) \delta \left(\partial_t a^2\right) + \frac{3}{2} \delta \left[ \left(\frac{1}{a^2 }\right) \cdot (\partial_t a^2) \right] \label{eq:8}.
\end{equation}
For details see  \ref{app:b}. 
Multiplying by $ \sqrt{-g}$, and taking the time derivative we get 
\begin{align}
\partial_0  \left(\sqrt{-g} B^0 \right)= & \partial_0 \left[\frac{3a^3}{2}  \left( \left(\frac{1}{a^2}\right) \delta \left(\partial_t a^2\right) + \frac{3}{2} \delta \left[ \left(\frac{1}{a^2 }\right) \cdot (\partial_t a^2) \right]  \right) \right] \nonumber \\ 
    = & \ \frac{3}{2} \partial_0 \left(   a \delta \left(2 a \dot a \right) +  a^3 \delta \left[ 2 \frac{\dot a }{a} \right]   \right) \label{cbt}.
\end{align}
The first term on the right hand side can be rewritten using Leibniz's rule as,
\begin{equation}
   a  \delta \left( a \dot a \right)=\delta   \left( a \cdot a \dot a \right)  -  a \dot a \delta \left( a \right) =  \delta \left( a^2 \dot a \right) -  a \dot a \delta \left( a \right).  
\end{equation}
Likewise, for the second term
\begin{align}
  a^3 \delta \left( \frac{\dot a }{a} \right)=  \delta \left( a^3 \frac{\dot a}{a} \right)  -  3 a\dot a \delta \left( a \right) =  \delta \left( a^2 \dot a \right)  -  3 a\dot a \delta \left( a \right)  ,
\end{align}
so that Eq. \eqref{cbt} takes the form 
\begin{align}\label{12}
    \partial_0  \left(\sqrt{-g} B^0 \right)= 6 \partial_0 \left(   \delta \left( a^2 \dot a \right) - 2 a \dot a \delta \left( a \right)  \right).
     \end{align}
In order to propose a new way to deal with the boundary terms, it is convenient to write the previous result as an exact variation. %For a variation in the action we need to take the total variation of a function,
% an exact variation.
{For this to be possible, the second term $2 a \dot a \delta \left( a \right)$ would need to be proportional to the first term $ \delta \left( a^2 \dot a \right)$, \textit{i.e.,}}%One possible way to rewrite the second term $2 a \dot a \delta \left( a \right)$ is if this term is proportional to the first term $ \delta \left( a^2 \dot a \right)$,
\begin{equation}
2 a \dot a \delta \left( a \right)   = A  \delta \left( a^2 \dot a \right) = A \left(  2 a \dot a \delta \left( a \right)  +  a^2 \delta \left( \dot a \right)\right),
\end{equation} 
where $A$ is the constant of proportionality. 
Assuming that $\dot{a}$ can be written as a function of $a(t)$ as is the case in the standard cosmological model, we can write, 
\begin{align}
2 a \dot a \delta \left( a \right)  
& = A \left(  2 a \dot a \delta \left( a \right)  +  a^2 \frac{\partial \dot a}{\partial a} \delta \left( a \right)\right)\nonumber \\
& = A \left(  2 a \dot a   +  a^2 \frac{\partial \dot a}{\partial a} \right) \delta \left( a \right),
\end{align} 
this implies that for $A\neq1$, in order to get the desired equality of the above equation
\begin{align}\label{15}
E a^2 \frac{\partial \dot a}{\partial a} = 2 a \dot a , 
\end{align}
where $E$ is another arbitrary  constant of proportionality. Putting all 
together,
\begin{align}
           2 a \dot a \delta \left( a \right)  
& = A\left(  1+ E\right)  2 a \dot a \delta \left( a \right), 
\end{align}
so $A= \frac{1}{1+E}$. Rearranging Eq. \eqref{15} and integrating we obtain
       \begin{equation}
         \dot a =  a^{2/E}. \label{srt} 
      \end{equation}
  %     this means that
  %    \begin{equation}
  %        a^2 \frac{\partial \dot a }{a }  = \frac{2}{E}   a^{2/E + 1},
  % \end{equation}and
  %     \begin{equation}
  %         2 a \dot a = 2 a^{2/E + 1 }.
  %     \end{equation}
As a consequence of the linearity  of Eq. \eqref{15} (considering $\dot a(a)$ as the independent variable), the general form of $\dot{a}$ that is compatible with the aforementioned conditions is a linear combination of powers of $a$, % as power series.
%The aforementioned conditions 
this guarantees that the RHS of Eq. \eqref{12} is a total variation. %\jf{Se podría quitar esto? Según yo no es esencial y así como está confunde un poco, lo quitamos o cambiamos la redacción? This means that the boundary term is,
%\begin{align}
%    \partial_0 (\sqrt{-g}B^0) & = 6 \partial_0 ((1-A)\delta \left(a^2 \dot a \right)) \nonumber \\
%    & = 6(1-A) \delta (2 a \dot a^2 + a^2 \ddot a )  \nonumber \\
%    & = 6\left(1 - \frac{1}{1+E}\right) \delta (2 a \dot a^2 + a^2 \ddot a ) \nonumber \\
%    & = \frac{6E}{1+E} \delta (2 a \dot a^2 + a^2 \ddot a ).
%\end{align}}

For the vanishing of the boundary term,
\begin{equation}
    2 a \dot a^2 + a^2 \ddot a = D_0,
\end{equation}
where $D_0$ is a constant. The solution yields
\begin{equation}
    a(t) = \frac{\left( \frac{3}{2}\right)^{1/3} \left( - c_1 + D_0^2 \left(t + c_2 \right)^2  \right)^{1/3}}{D_0^{1/3}} \label{Sola},
\end{equation}
where $c_1$ and $c_2$ are constants.

We must note that the case for $D_0=0$ should be treated separately,
\begin{equation}
    2 a \dot a^2 + a^2 \ddot a = 0,
\end{equation}
its solution is given by
\begin{equation}
 a(t) = (3 t - c_1)^{(1/3)} c_2 \label{Sola0} .
\end{equation}
{Notice that so far we have only considered the constraint equation. In the following, we incorporate this constraint to the variational principle in a cosmological setting. Specifically, we write the constraint as %Generally, the constraint equation is written as
\begin{equation}
  f(a,\dot a, \ddot a) =  \frac{1}{a^3} \left( 2 a \dot a^2 + a^2 \ddot a -D_0\right) = 0,  \label{int1}
\end{equation}
 %the vanishing boundary term it's successful by Eq.\eqref{int1} 
 which is a semi-holonomic  constraint, thus we obtain a well-posed Hamilton's principle in cosmology for GR, whose dynamical consequences are provided in the next section.}
%%%%%%%%%%%%%%%%%%%%%%%%%%%%%%%%%%%%
\section{Implementation of the boundary term in cosmology }\label{sec:3}
%%%%%%%%%%%%%%%%%%%%%%%%%%%%%%%%%%%%
When the variational principle includes \textit{auxiliary conditions} (such as boundary terms), it is useful to address them using the theory of Lagrange multipliers. These auxiliary conditions can take the form of \textit{holonomic, semi-holonomic}, or \textit{non-holonomic} constraint equations. In general, the Lagrange multiplier method is applicable in all these cases \cite{lanczos1986variational}.\\
The Lagrangian can be modified by adding the constraint equation with a Lagrange multiplier,
\begin{equation}
    L' = L + \lambda f,
\end{equation}
where the Lagrange multiplier may depend on time, $\lambda = \lambda(t)$ \cite{goldstein2002classical, courant89}. \\
Based on the previous analysis, we 
%Thus we arrive to the goal of the paper i.e., \textbf{to include boundary terms into the EFE} obtained via Hamilton's principle. The procedure mencioned above by itself predict an equation for the \textit{Hubble parameter} $H(t)$ squared where appear a term proportional to $1/a^6$, that's why we need to take in account the fluid \textit{kination} as a prediction due to the inclusion of boundary terms in GR. \\
%Kination was described as a fast rolling scalar field in many works \cite{Gouttenoire:2021jhk, Harigaya:2023pmw, Conlon:2024uob}. In this paper we only refer to it as a fluid proportial to $1/a^6$, related by an equation of state $\rho = \omega P$, with the value $\omega = 1$. 
%Despite the fact that originally the era of the Universe called kination was proposed after inflation and before that the Universe was dominated by radiation, some authors explore others places in the history of the universe where the term kination maybe may appear \cite{Gouttenoire:2021jhk}.
%Furthermore, kination may could appear when primordial gravitational waves can be produced \cite{Harigaya:2023pmw}, so the relics of kination may could be detected in future experiments.
obtain the  FLRW Lagrangian density that  incorporates the constraint that comes from the boundary term,
\begin{equation}
\sqrt{-g} \left(\mathcal{L} + \lambda(t) f\right) =  \frac{3}{\kappa N} a^2 \ddot a - 3 \frac{\dot N}{\kappa N^2} a^2 \dot a + \frac{3}{\kappa N} a \dot a^2 + a^3 N \lambda(t)  \frac{\left(2 a \dot a^2 + a^2 \ddot a -  D_0  \right)}{a^3}.
\end{equation}
The Euler-Lagrange equations for the variables of the system are,
\begin{align}
 f(a,\dot a, \ddot a) = \frac{1}{a^3} \left( 2 a \dot a^2 + a^2 \ddot a -D_0\right) &= 0, \label{eq:con} 
 \\
\lambda(t) \kappa \frac{\left( 2 a \dot a^2 + a^2 \ddot a  - D_0  \right)}{a^3} + 3 \frac{\dot a^2}{a^2}   &= 0 ,\label{nrho}
 \\
\kappa    \ddot \lambda(t)  + 3 \frac{\dot a^2}{a^2}  + 6 \frac{\ddot a }{a} &= 0 .\label{frw2}
\end{align}
where we set the gauge $N(t)= 1$ after calculating the Euler-Lagrange equations, and the first equation stands for the constraint equation due to the variation with respect to $\lambda(t)$.

Adding matter in the same way as it is done for the standard Friedmann equations, %Also from the standard cosmology comparing with the EFE, we know that the first Friedmann equation is
%\begin{equation}
%    3 \frac{\dot a^2 }{a^2 } = \kappa \rho_{eff} \label{frw1}
%\end{equation}while the second is
%\begin{equation}
%      \frac{\dot a^2 }{a^2 } + 2 \frac{\ddot a}{a} = - \kappa  P_{eff},
%\end{equation} (for the flat $k=0$ case)
with $\rho_{eff}$ and $P_{eff}$ %are
the \textit{effective} energy density and pressure, respectively, %.  Thus 
we can write the evolution equations as
\begin{align}
     \lambda(t) \kappa \frac{\left(   2 a \dot a^2 + a^2 \ddot a - D_0 \right)}{a^3} + 3 \frac{\dot a^2}{a^2}   &= \kappa \rho_{eff} ,\label{nrho1}
\\
    \kappa    \ddot \lambda(t)  + 3 \frac{\dot a^2}{a^2}  + 6 \frac{\ddot a }{a} &= - 3 \kappa P_{eff} .\label{frw22}
\end{align}
{Summarizing, Eqs.~\eqref{eq:con}, \eqref{nrho1}, and \eqref{frw22} are the constrained dynamical equations for a FRW universe with matter described by perfect fluids.  } 
%Thus we add the matters terms into the Friedmann equation from the EFE.
%%%%%%%%%%%%%%%%%%%%%%%%%%%%%%%%%%%%%%%%%%%%%%%%%%%%%%%%%%%%%%%
\subsection{Consequences of the boundary term.}
%%%%%%%%%%%%%%%%%%%%%%%%%%%%T
To complete the analysis of the BT into the EFE, we solve the Friedmann equations in this context. The term in parenthesis in Eq.~\eqref{nrho1} is the constraint itself, so it is equal to zero. By using Eq. \eqref{Sola} into Eq. \eqref{nrho1} we get 
\begin{equation}
    \frac{2}{3} \frac{D_0}{a^3} + \frac{c_1}{a^6} = \frac{\kappa}{3}  \rho_{eff}. \label{frw}
\end{equation}
%from Eq.\eqref{frw1} this is 
Here we identify an expression for the squared Hubble parameter, $H(t)^2$, which includes a term proportional to $1/a^6$% (the evolution factor)
. %i.e. for the $\rho_{eff}$ 
{From the equation of state and from the continuity equation,
\begin{equation}
   \rho = \rho_0 a^{-3(1+\omega)},
\end{equation}
so the term proportional to $1/a^6$ is linked to a stiff matter fluid  characterized by $\omega = 1$. This fluid has the particularity that the speed of sound is equal to the speed of light, so it is the maximum speed of sound allowed by causality \cite{Zeldovich:1961sbr, Joyce:1996cp}.}
In this way we can write
\begin{align}
    3 \frac{\dot a^2}{a^2} = \kappa  \rho_{eff}=  \kappa \left( \frac{\rho_{0d}}{a^3}  + \frac{\rho_{0s}}{a^6} \right),  \label {Fried2}
\end{align}
{where we have rewritten $D_0$ and $c_1$ in terms of the present day densities of dust and stiff matter, respectively, } %by comparison of the eq. \eqref{frw1} and the above equation we see that, 
$D_0 = {\kappa \rho_{0d} }/{2}$ for dust, and $ c_1 = {\kappa \rho_{0s} }/{3}$ for stiff.
This dependence predicts the existence of a stiff matter fluid as an emergent consequence of including a boundary term in GR.

\begin{figure}
    \centering
\includegraphics[scale=.6]{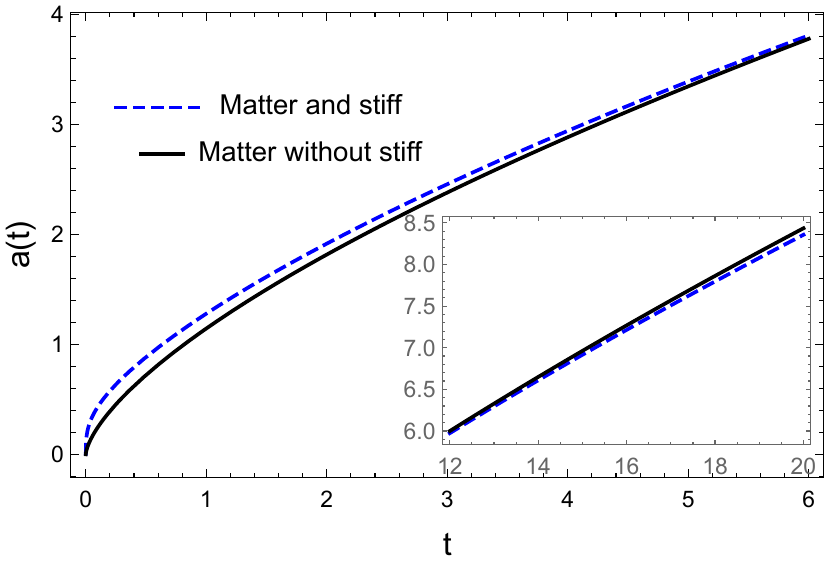}
    \caption{From Eq.\eqref{Sola} we set $c_2 = \sqrt{c_1}/D_0$. In this qualitative analysis, $3H_0^2 = 1$. The dashed blue line corresponds to  $c_1=0.05$, $D_0= 0.95$, while the black line is for $c_1=0, D_0 =1$.}
    \label{MS0}
\end{figure}
In Fig.~\ref{MS0} we show the different behavior of the scale factor $a(t)$, Eq.~\eqref{Sola}, for cases with and without stiff matter. At early times we have a slightly faster volume expansion for the case of dust and stiff matter than for dust without stiff matter. At later times the scale factor grows faster in the case of dust without stiff matter. This happens because the density of dust, which dominates at later times, is higher when there is no stiff matter. 

For the second constrained Friedmann  equation, substituting  Eq.~\eqref{Sola} into Eq.~\eqref{frw22} we obtain
\begin{align}
   \left(-4 \rho_{0s} + 3 \kappa \rho_{0d}^2\left(t+c_2\right)^2\right)^{1/3} \ddot \lambda(t) =0 ,
\end{align}
with solution
\begin{align}
    \lambda(t) & = c_3+ c_4 t \label{lambdasol2},
\end{align}
where $c_3$ and $c_4$ are integration constants that determine the initial conditions for $\lambda(t)$. We note that these constants do not appear in the solution Eq.~\eqref{Sola} for the scale factor, \textit{i.e.}, the evolution of the scale factor is not sensitive to the initial conditions of the Lagrange multiplier. 
%parameters can be chosen such that they do not affect the evolution equation.\\
%And for the time derivative, with initial conditions $\dot \lambda (t_0)$,
%\begin{align}
%  \dot  \lambda(t_0) & =  c_4  \label{dotlambdasol}
%\end{align}
%again the free parameter $c_4$ does not affect the evolution equations.
%%%%%%%%%%%%%%%%%%%%%%%%%%%%%%%%%%%%%%%
\subsection{Relaxing the boundary term: the cosmological constant}
%%%%%%%%%%%%%%%%%%%%%%%%%%%%%%%%%%%%%
A caveat of the previous results is that they do not contain a term capable of reproducing the accelerated expansion of the Universe indicated by observations \cite{Riess1998,Perlmutter1999}. In order to achieve such a behavior, we modify our considerations by allowing for contributions to the constraint equation that become negligible at the boundary as the Universe expands. For instance, we have:\footnote{Our criteria for deciding which terms may be added comes from noting that in the variation of the GR action, Eq.~\eqref{GRvar}, the terms corresponding to the EFE are multiplied by $\sqrt{-g}$ ($a^3$ for the FRW metric). Therefore, we allow for terms such that $\partial_\sigma(\sqrt{-g} B^\sigma)/\sqrt{-g}\to0$. This condition is fulfilled by $F_1 a^3$ as $\delta(a^3)/a^3 \sim \delta a/a$ which vanishes as $a$ gets larger.}
\begin{equation}
    2 a \dot a^2 + a^2 \ddot a = D_0 + F_1 a^3.
\end{equation}
The solution to this equation is
\begin{align}
    a(t) & = \frac{1}{F_1^{1/3} 2^{1/3}} e^{\frac{1}{3}\left(-\sqrt{3}\sqrt{F_1}\left(t+ c_2\right)\right)} \nonumber \\
    & \times \left(e^{2\sqrt{3}\sqrt{F_1} \left(t+ c_2\right)}-2e^{\sqrt{3}\sqrt{F_1} \left(t+ c_2\right)} D_0+ D_0^2 - 3 F_1 c_1\right)^{1/3}, \label{SolL}
\end{align}
where $c_1, c_2$ are constants.
The modified GR Lagrangian density is given by:
\begin{align}
   \sqrt{-g} \left(\mathcal{L} + \lambda(t) f \right) & =  \frac{3}{\kappa N} a^2 \ddot a - 3 \frac{\dot N}{\kappa N^2} a^2 \dot a + \frac{3}{\kappa N} a \dot a^2 
   \nonumber \\
   &  + a^3 N \lambda(t)  \frac{\left(2 a \dot a^2 + a^2 \ddot a -  D_0 - F_1 a^3 \right)}{a^3}.
\end{align} 
The Euler-Lagrange equations are
\begin{align}
    \lambda(t) \kappa \frac{\left( - D_0 - F_1 a^3 + 2 a \dot a^2 + a^2 \ddot a  \right)}{a^3} + 3 \frac{\dot a^2}{a^2}   &= k\rho_{eff} ,\label{nrho2}\\
    \kappa \left( - 3 F_1\lambda(t) +   \ddot \lambda(t) \right) + 3 \frac{\dot a^2}{a^2}  + 6 \frac{\ddot a }{a}& = -kP_{eff},
\end{align}
Using Eq.~\eqref{SolL} in Eq.~\eqref{nrho2} we deduce
\begin{align}
      3 \frac{\dot a^2}{a^2} &  = \frac{2D_0}{a^3} + F_1 + \frac{3 c_1}{a^6} \\
   & \equiv  \kappa \left( \frac{\rho_{0d}}{a^3} + \rho_{0\Lambda } + \frac{\rho_{0s}}{a^6} \right),  \label{Fried2}
\end{align}
where we identify  $D_0 = {\kappa \rho_{0d} }/{2}$, $ c_1 =   {\kappa \rho_{0s} }/{3}$ and $F_1 = \kappa \rho_{0\Lambda}$ for dust, stiff matter and cosmological constant, respectively.

\begin{figure}
    \centering
    \includegraphics[scale=.6]{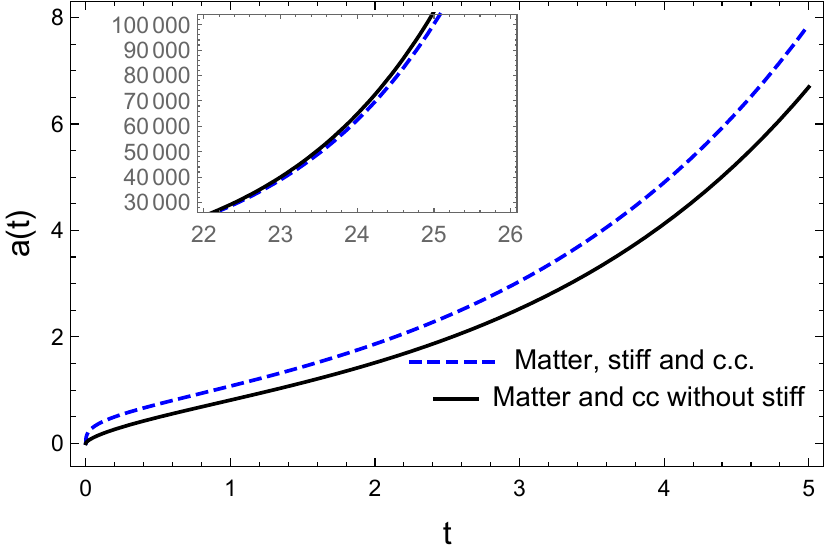}
    \caption{From the Eq.\eqref{SolL} we set $c_2 = \log{\left(D_0+\sqrt{3 F_1 c_1}\right)}/\sqrt{3 F_1 }$. Again for this qualitative analysis, $3H_0^2 = 1$.  The dashed blue line corresponds to $c_1=0 .03$, $D_0=0.3$ and $F_1= 0.67$, while the black line is for $D_0=0.3$ and $F_1= 0.7$.}
    \label{MCCS0}
\end{figure}
In Fig. \ref{MCCS0} we see how the presence of stiff matter makes the scale factor grow faster at early times, notably at large times the scale factor  grows faster when there is no stiff matter.

Using the previous results in the second constrained Friedmann equation we get
\begin{align}
   3 k \rho_{0\Lambda} \lambda(t)- \ddot\lambda(t)= 0     .
\end{align}
Integrating to obtain the solution for $\lambda(t)$ we obtain
  \begin{align}
    \lambda(t) & = e^{\sqrt{3\kappa \rho_{0\Lambda}}  t} c_3+ e^{-\sqrt{3\kappa \rho_{0\Lambda}}  t}c_4.
    \end{align}
    Again the free parameters $c_3, c_4$ determine the initial conditions  $\lambda(t_0), \dot \lambda(t_0)$.
    
Finally, let us analyze the evolution of the Hubble parameter in this cosmological model.
\begin{figure}
    \centering
    \includegraphics[scale=.615]{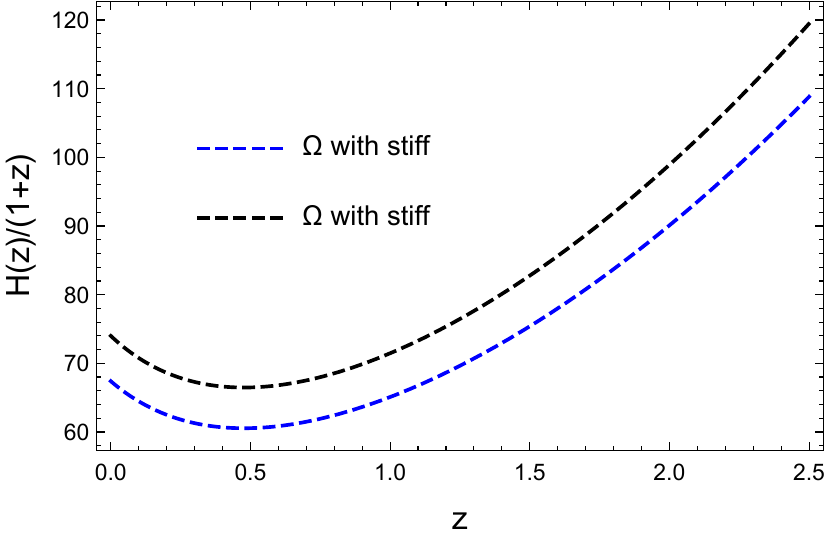}
    \caption{The dashed blue line and the black line are set at $H_0 = 67.4$ km/s/Mpc, while the dashed red line is set at $H_0 = 74.03$ km/s/Mpc.}
    \label{Omega}
\end{figure}
From observations, two differente values of the Hubble parameter at present time have been reported: $H_0 = 67.4$ km/s/Mpc \cite{Planck2018VI} and $H_0 = 74.03$ km/s/Mpc \cite{Riess2019}. These values differ by more than the accepted margin of error associated with the measurements. As illustrated in Fig.\eqref{Omega}, this so-called Hubble tension is not resolved by the introduction of a stiff fluid component. This issue has been notably addressed in Ref.~\refcite{Oyvind:2024axi} .
%These values are away from each other more than the accept error of the measurements. This Hubble tension as seen in the Fig.\eqref{Omega} is not solve by mean of the stiff fluid. This fact was notably worked in ref \cite{Oyvind:2024axi} .

%%%%%%%%%%%%%%%%%%%%%%%
\subsubsection{Numerical solution including radiation}
%%%%%%%%%%%%%%%%%%%%%%%
Under the same considerations used to introduce the cosmological constant, we introduce the term corresponding to radiation into the constraint equation as
\begin{equation}
    2 a \dot a^2 + a^2 \ddot a = D_0 + F_1 a^3 + \frac{R_2}{a} .
\end{equation}
  \begin{figure}
        \centering
        \includegraphics[scale=.6]{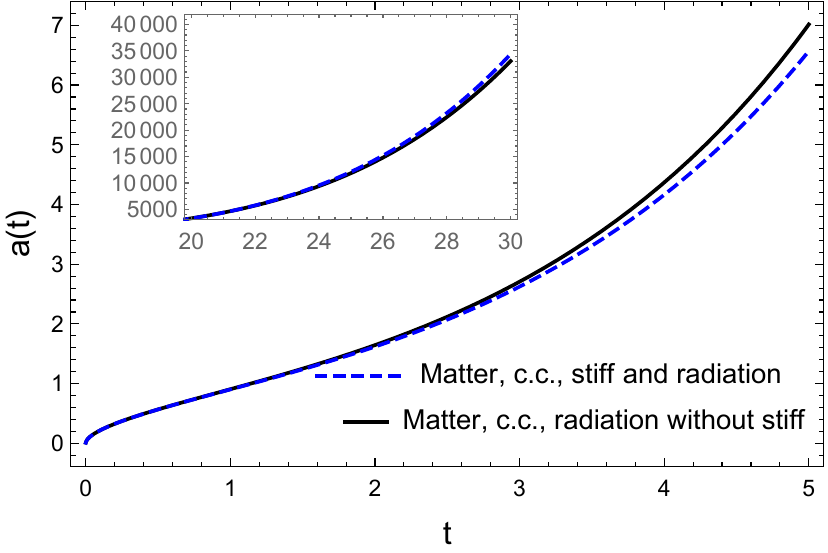}
        \caption{ In this qualitative analysis, $3H_0^2 = 1$. The dashed blue line corresponds to $R_2= 0.03$, $D_0=0.3$ and $F_1= 0.62$, while the black line is for $R_2= 0.03$, $D_0=0.3$ and $F_1= 0.67$.}
        \label{MCCRS01}
    \end{figure}
We solve this equation numerically. 
Once again, the modified Lagrangian density becomes
\begin{align}
   \sqrt{-g} \left(\mathcal{L} + \lambda(t) f \right)  =&  \frac{3}{\kappa N} a^2 \ddot a - 3 \frac{\dot N}{\kappa N^2} a^2 \dot a + \frac{3}{\kappa N} a \dot a^2 
   \nonumber \\
     &+ a^3 N \lambda(t)  \frac{\left(2 a \dot a^2 + a^2 \ddot a -  D_0 - F_1 a^3 - \frac{R_2}{a}\right)}{a^3}.
\end{align} 
The Euler-Lagrange equations of motion are
\begin{align}
    \lambda(t) \kappa \frac{\left( - D_0 - F_1 a^3 - \frac{R_2}{a}+ 2 a \dot a^2 + a^2 \ddot a  \right)}{a^3} + 3 \frac{\dot a^2}{a^2}   = k\rho_{eff} ,\label{nrho3}
\end{align}
\begin{align}
    \kappa \left( - 3 F_1\lambda(t) +   \ddot \lambda(t) \right) + \kappa \frac{\lambda(t)R_2}{a^3} + 3 \frac{\dot a^2}{a^2}  + 6 \frac{\ddot a }{a} = -kP_{eff} .\label{apres}
\end{align}
In Fig. \ref{MCCRS01} we see how the scale factor $a(t)$ is modified by the inclusion of the stiff matter in early times of the Universe, and at later times it grows faster when there is no stiff matter. Again, as we have a second order differential equation for $\lambda(t)$, we can choose the two free parameters in a convenient way to determine the initial conditions of $\lambda(t)$ without affecting the evolution equations directly.

%%%%%%%%%%%%%%%%%%%%%%%%%%%%%%%%%%%%
\section{Conclusion and discussions}\label{sec:4}
%%%%%%%%%%%%%%%%%%%%%%%%%%%%%%%%%%%%
In this work, we presented a novel formulation of the Einstein field equations for a cosmological FLRW setting, integrating an additional constraint to enforce specific boundary conditions. We find that the standard Friedmann equations are preserved, but some of the matter content arises as a contribution from the constraint that enforces the vanishing of the boundary term.  Notably, while the concept of a stiff matter fluid has been theorized in previous studies, in this model, it arises naturally as a prediction, highlighting the model’s potential to contribute new perspectives on early Universe cosmology.

We remark some aspects and  possible research directions that follow from our results: 
\begin{itemize}
    \item \textit{Early Universe dynamics}.  As we have seen in the three cases, the continuity equation together with the first Friedmann equation imply the usual second Friedmann equation, therefore in the modified second Friedmann equation the only terms that survive are those that multiply $\ddot \lambda(t)$, which leads to a linear solution for $\lambda(t)$ in the first case (dust) and to an exponential solution in the second case (dust  and cosmological constant). When radiation is considered, the solution for $\lambda(t)$ is numerical. An important fact is that the stiff matter fluid cannot resolve the Hubble tension, but in early times it changes the rate of expansion of the Universe.
    \item \textit{Alternative cosmological metrics}. A natural extension of this model would be to explore its application to other cosmological metrics, such as Kantowski-Sachs or Bianchi models, as well as to non-asymptotically flat metrics and astrophysical configurations. Such explorations could reveal additional phenomena or constraints unique to each metric.
    \item \textit{Boundary terms and additional matter fields}. For the matter boundary term to be non-vanishing, the vector density $G^\sigma$ must depend on the derivatives of the metric tensor.  This condition underscores the role of matter fields that couple dynamically to the metric derivatives, which could result in distinctive boundary behaviors. The inclusion of non-minimally coupled additional matter fields, such as scalar or vector fields, may lead to non-trivial contributions to the boundary terms. This opens the possibility of a richer interaction between the matter sector and the boundaries, potentially affecting the evolution of the universe in significant ways. For instance, it would be interesting to explore non-minimally coupled scalar fields that in some limit behave as kination, since this field appears naturally in our setup.
    %.example as an scalar field fluid, despite the fact that originally the era of the Universe called kination was proposed after inflation and before that the Universe was dominated by radiation, some authors explore others places in the history of the universe where the term kination maybe may appear \cite{Gouttenoire:2021jhk}. Furthermore, kination may could appear when primordial gravitational waves can be produced \cite{Harigaya:2023pmw}, or in cosmic strings \cite{Conlon:2024uob}, so the relics of kination may could be detected in future experiments. 
    \item \textit{Covariant generalization and initial value problems}. A covariant extension of this approach would allow for a more comprehensive treatment of boundary terms across different spacetime configurations. Revisiting classical results, such as initial value problems in general relativity, under this framework could provide new insights, particularly regarding the role and impact of boundary conditions. 
\end{itemize}

In particular the analysis of the non-minimal couplings and the covariant formulations of our approach are critical for a better understanding of the effect of boundary terms in cosmology and gravitation. Results along these lines will be reported elsewhere.

\section*{Acknowledgments}

A.C. is supported by CONAHCyT grant 847964. JC acknowledges support from DCF-320821.

\appendix
\section{Deduction of EFE from Hamilton's principle}
Starting from the action of General Relativity:
\begin{equation}
           S_{RG}=  S_{EH} + S_M =  \int \left(\frac{R}{2\kappa} + \mathcal{L_M}\right) \sqrt{-g} d^4x.
       \end{equation}  
We obtain the equations of motion using the principle of least action,
    \begin{align}
        \delta  S_{RG} & = \delta \int \left(\frac{R}{2\kappa} + \mathcal{L_M}\right) \sqrt{-g} d^4x = \int \delta \left[ \left(\frac{R}{2\kappa} + \mathcal{L_M}\right) \sqrt{-g} \right] d^4x \nonumber \\
        & =  \int \left[\delta \left(\frac{R}{2\kappa}\right)\sqrt{-g} + \frac{R}{2\kappa} \delta\sqrt{-g} + \delta (\sqrt{-g} \mathcal{L_M})\right]d^4x \nonumber \\
        &  = \int \left[\frac{g^{\alpha \beta} \delta R_{\alpha \beta}  + \delta g^{\alpha \beta} R_{\alpha \beta} }{2\kappa}\sqrt{-g} + \frac{R_{\alpha \beta} g^{\alpha \beta}}{2\kappa} \delta\sqrt{-g} + \delta (\sqrt{-g} \mathcal{L_M})\right]d^4x,
         \end{align}
where
         \begin{align}
            \delta g^{\alpha \beta} R_{\alpha \beta} \sqrt{-g} + R_{\alpha \beta} g^{\alpha \beta} \delta \sqrt{-g} &=  \sqrt{-g} (\delta g^{\alpha \beta}  -  \frac{1}{2} g^{\alpha \beta} g_{\rho \sigma} \delta g^{\rho \sigma}) R_{\alpha \beta}\nonumber\\
            &=\sqrt{-g} \left( R_{\alpha \beta} - \frac{1}{2} R  g_{\alpha \beta}\right) \delta g^{\alpha \beta}.
        \end{align}
and 
\begin{align}
   & g^{\alpha \beta} \delta R_{\alpha \beta} = \nabla_\sigma ( g^{\alpha \beta} \delta \Gamma^\sigma_{\beta \alpha} - g^{\alpha \sigma}\delta \Gamma^\beta_{\beta \alpha}) \equiv  \nabla_\sigma B^\sigma, \nonumber \\
    &    \sqrt{-g} \nabla_\sigma B^\sigma = \nabla_\sigma( \sqrt{-g} B^\sigma) = \partial_\sigma( \sqrt{-g} B^\sigma).
\end{align}
Assuming that the Lagrangian of matter depends on the metric and its first derivatives, the variation $\delta \left(\sqrt{-g} \mathcal{L_M}\right)$ takes the form 
      \begin{equation}
       \delta \left(\sqrt{-g} \mathcal{L_M}\right) = \frac{\partial \left(\sqrt{-g} \mathcal{L_M}\right)}{\partial g^{\alpha \beta}} \delta g^{\alpha \beta} +  \frac{\partial \left(\sqrt{-g} \mathcal{L_M}\right)}{\partial (\partial_\sigma g^{\alpha \beta})} \partial_\sigma \delta g^{\alpha \beta}.\label{eq:varlm}
    \end{equation}
%Sabemos que las partículas relativistas se asumen como partículas libres (solo la parte cinética contribuye). Los cuadrivectorees velocidad provienen de una derivada parcial. Introduciendo una derivada covariante reemplazando a la derivada parcial obtenemos un lagrangiano más acorde un espacio curvo. \\ \\
A vector $G^\sigma$ is defined such that,
\begin{equation} G^\sigma = 
    \frac{\partial \left(\sqrt{-g} \mathcal{L_M}\right)}{\partial (\partial_\sigma g^{\alpha \beta})} \delta g^{\alpha \beta}.
\end{equation} 
Taking the divergence of this vector we get,
\begin{equation}
\partial_\sigma G^\sigma = \partial_\sigma \frac{\partial \left(\sqrt{-g} \mathcal{L_M}\right)}{\partial (\partial_\sigma g^{\alpha \beta})} \delta g^{\alpha \beta} +  \frac{\partial \left(\sqrt{-g} \mathcal{L_M}\right)}{\partial (\partial_\sigma g^{\alpha \beta})} \partial_\sigma \delta g^{\alpha \beta},
\end{equation}
rearranging,
\begin{equation}
     \frac{\partial (\sqrt{-g} \mathcal{L_M})}{\partial (\partial_\sigma g^{\alpha \beta})} \partial_\sigma \delta g^{\alpha \beta} = \partial_\sigma G^\sigma - \partial_\sigma  \frac{\partial (\sqrt{-g} \mathcal{L_M})}{\partial (\partial_\sigma g^{\alpha \beta})} \delta g^{\alpha \beta}.
\end{equation} 
By introducing this into the ec.~\eqref{eq:varlm},
     \begin{equation}
        \delta \left(\sqrt{-g} \mathcal{L_M}\right) = \frac{\partial (\sqrt{-g} \mathcal{L_M})}{\partial g^{\alpha \beta}} \delta g^{\alpha \beta} + \partial_\sigma G^\sigma - \partial_\sigma  \frac{\partial (\sqrt{-g} \mathcal{L_M})}{\partial (\partial_\sigma g^{\alpha \beta})} \delta g^{\alpha \beta}.
    \end{equation}
    Therefore,
    \begin{align}
    \int \delta \left(\sqrt{-g} \mathcal{L_M}\right)d^4x &  =    \int \left(\frac{\partial (\sqrt{-g} \mathcal{L_M})}{\partial g^{\alpha \beta}} \delta g^{\alpha \beta} - \partial_\sigma  \frac{\partial (\sqrt{-g} \mathcal{L_M})}{\partial (\partial_\sigma g^{\alpha \beta})} \delta g^{\alpha \beta}\right) d^4x \\
    & = -\frac12 \int \sqrt{-g} T_{\alpha\beta}\delta g^{\alpha\beta}d^4x + \int \partial_\sigma G^\sigma d^4x,
    \end{align}
     where:
     \begin{equation}
        T_{\alpha\beta} = \frac{-2}{\sqrt{-g}}\left(\frac{\partial (\sqrt{-g} \mathcal{L_M})}{\partial g^{\alpha \beta}}  - \partial_\sigma  \frac{\partial (\sqrt{-g} \mathcal{L_M})}{\partial (\partial_\sigma g^{\alpha \beta})} \right),
     \end{equation} 
     is defined as the \textbf{energy-momentum tensor.}\\Returning to the variation of the total action of General Relativity we obtain the Eq.~\eqref{GRvar} presented in the main text.
%%%%%%%%%%%%%%%%%%%%%%%%%%%%%%%%%%%%
\section{Value of the boundary term $B^\sigma$}\label{app:b}
%%%%%%%%%%%%%%%%%%%%%%%%%%%%%%%%%%%%
Here we derive Eq.~\eqref{eq:8} by evaluating $B^\sigma$ in the FLRW spacetime with metric
    \begin{equation} 
    g_{\alpha \beta} = \begin{pmatrix} - N^2 (t)& 0 & 0 & 0\\ 0 &  \frac{a^2(t)}{1 - kr^2} & 0 & 0 \\ 0 & 0 &  a^2(t) r^2  & 0  \\ 0 & 0 & 0 &  a^2(t) r^2 \sin^2{\theta} \end{pmatrix}.
     \end{equation}  
Since the field equations are compatible with $N=1$, we take the same assumption for evaluating $B^\sigma$. By analyzing the vector $B^\sigma = g^{\alpha \beta} \delta \Gamma^\sigma_{\beta \alpha} - g^{\alpha \sigma}\delta \Gamma^\beta_{\beta \alpha}$, where
     
     \begin{equation} \delta \Gamma^\sigma_{\beta\alpha} =  \frac{1}{2}  \delta g^{\sigma\lambda} (\partial_\alpha g_{\lambda\beta} + \partial_\beta g_{\lambda \alpha} - \partial_\lambda g_{\alpha\beta}) +
     \frac{1}{2} g^{\sigma\lambda} (\partial_\alpha (\delta g_{\lambda\beta}) + \partial_\beta (\delta g_{\lambda \alpha}) - \partial_\lambda (\delta g_{\alpha\beta})),
\end{equation}
%  \begin{equation} \delta \Gamma^\beta_{\beta\alpha} =  \frac{1}{2}  \delta g^{\beta\lambda} (\partial_\alpha g_{\lambda\beta} + \partial_\beta g_{\lambda \alpha} - \partial_\lambda g_{\alpha\beta}) +
%      \frac{1}{2} g^{\beta\lambda} (\partial_\alpha (\delta g_{\lambda\beta}) + \partial_\beta (\delta g_{\lambda \alpha}) - \partial_\lambda (\delta g_{\alpha\beta})),
% \end{equation} 
So the components of the vector $B^\sigma$ are,
\begingroup
\allowdisplaybreaks
 \begin{align}
    B^\sigma  = & \Biggl[ g^{\alpha \beta} \delta \Gamma^0_{\beta \alpha} - g^{\alpha 0}\delta \Gamma^\beta_{\beta \alpha} , \ g^{\alpha \beta} \delta \Gamma^1_{\beta \alpha} - g^{\alpha 1}\delta \Gamma^\beta_{\beta \alpha} , \ g^{\alpha \beta} \delta \Gamma^2_{\beta \alpha} - g^{\alpha 2}\delta \Gamma^\beta_{\beta \alpha},
    \nonumber \\
    & g^{\alpha \beta} \delta \Gamma^3_{\beta \alpha} - g^{\alpha 3}\delta \Gamma^\beta_{\beta \alpha}  \Biggl] \nonumber \\
    & = \Biggl\{ \Biggl[ \frac{1}{2} \left(\frac{1 - kr^2}{a^2}\right) \delta \left(\frac{\partial_t a^2}{1 - kr^2}\right) +  \frac{1}{2}   \left(\frac{1}{a^2 r^2}\right)  \delta (\partial_t a^2 r^2) \nonumber \\
    & +  \frac{1}{2} \left(\frac{1}{a^2 r^2 \sin^2{\theta}}\right)  \delta  (\partial_t a^2 r^2 \sin^2{\theta})
   + \frac{1}{2} \delta \left[ \left(\frac{1 - kr^2}{a^2}\right) \cdot \left(\frac{\partial_t a^2}{1 - kr^2}\right) \right] 
    \nonumber \\
   &  + \frac{1}{2} \delta \left[ \left(\frac{1}{a^2 r^2}\right) \cdot (\partial_t a^2 r^2) \right] + \frac{1}{2} \delta \left[ \left(\frac{1}{a^2 r^2 \sin^2{\theta}}\right) \cdot (\partial_t a^2 r^2 \sin^2{\theta}) \right] \Biggl], \nonumber \\
&   \Biggl[ - \frac{1}{2} \left(\frac{1}{a^2 r^2}\right) \delta \left[ \left(\frac{1 - kr^2}{a^2}\right) \cdot \left(\partial_r a^2 r^2\right) \right] \nonumber \\
& - \frac{1}{2} \left(\frac{1}{a^2 r^2 \sin^2{\theta}}\right) \delta \left[ \left(\frac{1 - kr^2}{a^2}\right) \cdot \left(\partial_r a^2 r^2 \sin^2{\theta} \right) \right]  \Biggl], \nonumber \\
&  \Biggl[- \frac{1}{2} \left(\frac{1}{a^2 r^2 \sin^2{\theta}}\right) \delta \left[ \left(\frac{1}{a^2 r^2}\right) \cdot \left(\partial_\theta a^2 r^2 \sin^2{\theta} \right) \right] \nonumber \\ 
& - \frac{1}{2} \left(\frac{1}{a^2 r^2}\right) \delta \left[ \left(\frac{1}{a^2 r^2 \sin^2{\theta}}\right) \cdot \left(\partial_\theta a^2 r^2 \sin^2{\theta} \right) \right] \Biggl],0\Biggl\}.
\end{align}
\endgroup
We consider a boundary at some $r = constant$, from spherical symmetry we choose $\theta = \frac{\pi}{2}$, and we focus on the spatially flat case $k=0$. Under these assumptions, the only non-vanishing component of $B^\sigma$ is
\begin{align}
      B^0 = & \ \frac{1}{2} \left(\frac{1}{a^2}\right) \delta  (\partial_t a^2) +  \frac{1}{2}   \left(\frac{1}{a^2}\right)  \delta (\partial_t a^2) +  \frac{1}{2} \left(\frac{1}{a^2}\right)  \delta (\partial_t a^2) 
   \nonumber \\
   &   + \frac{1}{2} \delta \left[ \left(\frac{1}{a^2}\right) \cdot \left(\partial_t a^2\right) \right] + \frac{1}{2} \delta \left[ \left(\frac{1}{a^2}\right) \cdot (\partial_t a^2) \right] + \frac{1}{2} \delta \left[ \left(\frac{1}{a^2 }\right) \cdot (\partial_t a^2) \right]    \nonumber \\
     = & \ \frac{3}{2} \left(\frac{1}{a^2}\right) \delta \left(\partial_t a^2\right) + \frac{3}{2} \delta \left[ \left(\frac{1}{a^2 }\right) \cdot (\partial_t a^2) \right].
\end{align}
which is Eq.~\eqref{eq:8}.

\bibliographystyle{unsrt}
\bibliography{ref}

\begin{thebibliography}{10}

\bibitem{Krishnan:2016mcj}
Chethan Krishnan and Avinash Raju.
\newblock A neumann boundary term for gravity.
\newblock {\em Journal of High Energy Physics}, 8, 2017.
\newblock arXiv:1608.06223.

\bibitem{Izumi:2023rwh}
Keisuke Izumi, Keigo Shimada, Kyosuke Tomonari, and Masahide Yamaguchi.
\newblock Boundary conditions for constraint systems in variational principle.
\newblock {\em Journal of High Energy Physics}, 9, 2023.

\bibitem{DeHaro:2021gdv}
Sebastian De~Haro.
\newblock Noether's theorems and energy in general relativity.
\newblock {\em Foundations of Physics}, 66, 2021.

\bibitem{Hilbert1915}
D.~Hilbert.
\newblock Die grundlagen der physik.
\newblock {\em Nachrichten von der Gesellschaft der Wissenschaften zu Göttingen, Mathematisch-Physikalische Klasse}, pages 395--407, 1915.

\bibitem{lanczos1986variational}
C.~Lanczos.
\newblock {\em The Variational Principles of Mechanics}.
\newblock Dover Books On Physics. Dover Publications, 1986.

\bibitem{goldstein2002classical}
H.~Goldstein, C.P. Poole, and J.L. Safko.
\newblock {\em Classical Mechanics}.
\newblock Addison-Wesley Series in Physics. Addison-Wesley, 3rd edition, 2002.

\bibitem{gron2007einstein}
{\O}.~Gr{\o}n and S.~Hervik.
\newblock {\em Einstein's General Theory of Relativity: With Modern Applications in Cosmology}.
\newblock Springer New York, 2007.

\bibitem{Carroll:2004st}
Sean~M. Carroll.
\newblock {\em {Spacetime and Geometry}: {An Introduction to General Relativity}}.
\newblock Cambridge University Press, 7 2019.

\bibitem{Parattu:2016trq}
Krishnamohan Parattu, Sumanta Chakraborty, and T.~Padmanabhan.
\newblock Variational principle for gravity with null and non-null boundaries: A unified boundary counter-term.
\newblock {\em Journal of High Energy Physics}, 5, 2016.

\bibitem{Chakraborty:2016yna}
Sumanta Chakraborty.
\newblock {Boundary Terms of the Einstein\textendash{}Hilbert Action}.
\newblock {\em Fundam. Theor. Phys.}, 187:43--59, 2017.

\bibitem{Zeldovich:1961sbr}
Ya.~B. Zel'dovich.
\newblock The equation of state at ultrahigh densities and its relativistic limitations.
\newblock {\em Soviet Physics JETP}, 7:1163--1167, 1961.

\bibitem{Joyce:1996cp}
Michael Joyce.
\newblock {Electroweak Baryogenesis and the Expansion Rate of the Universe}.
\newblock {\em Phys. Rev. D}, 55:1875--1878, 1997.

\bibitem{hadamard1902problemes}
Jacques Hadamard.
\newblock Sur les probl{\`e}mes aux d{\'e}riv{\'e}es partielles et leur signification physique.
\newblock {\em Princeton university bulletin}, pages 49--52, 1902.

\bibitem{York:1986lje}
James York.
\newblock {Boundary terms in the action principles of general relativity}.
\newblock {\em Found. Phys.}, 16:249--257, 1986.

\bibitem{Gibbons:1976ue}
G.~W. Gibbons and S.~W. Hawking.
\newblock {Action Integrals and Partition Functions in Quantum Gravity}.
\newblock {\em Phys. Rev. D}, 15:2752--2756, 1977.

\bibitem{chrusciel2004einstein}
P.T. Chru{\'s}ciel and H.F. Friedrich.
\newblock {\em The Einstein Equations and the Large Scale Behavior of Gravitational Fields: 50 Years of the Cauchy Problem in General Relativity}.
\newblock Birkh{\"a}user, Basel, 2004.

\bibitem{Karp:2009tq}
Lavi Karp.
\newblock On the well-posedness of the vacuum einstein's equations, 2009.
\newblock arXiv:0906.0137.

\bibitem{Isenberg:2013iva}
James Isenberg.
\newblock The initial value problem in general relativity.
\newblock In {\em Springer Handbook of Spacetime}, pages 303--321. Springer, 2014.

\bibitem{Peacock:1999ye}
J.~A. Peacock.
\newblock {\em {Cosmological physics}}.
\newblock Cambridge university press, 1999.

\bibitem{courant89}
Richard Courant and David Hilbert.
\newblock {\em Methods of Mathematical Physics}, volume~1.
\newblock Wiley, New York, 1989.

\bibitem{Riess1998}
A.~G. Riess, W.~H. Press, and R.~P. Kirshner.
\newblock Type ia supernovae and the accelerating universe: A new approach.
\newblock {\em The Astronomical Journal}, 116(3):1009--1026, 1998.

\bibitem{Perlmutter1999}
Saul Perlmutter, Brian~P. Schmidt, Adam~G. Riess, et~al.
\newblock Supernovae, dark energy, and the accelerating universe.
\newblock {\em The Astrophysical Journal}, 517(2):565--586, 1999.

\bibitem{Planck2018VI}
Planck Collaboration.
\newblock Planck 2018 results. vi. cosmological parameters.
\newblock {\em Astronomy \& Astrophysics}, 641:A6, 2020.
\newblock Erratum: \url{https://doi.org/10.1051/0004-6361/201833910e}.

\bibitem{Riess2019}
Adam~G. Riess, Stefano Casertano, Wenlong Yuan, Lucas~M. Macri, and Dan Scolnic.
\newblock Large magellanic cloud cepheid standards provide a 1\ foundation for the determination of the hubble constant and stronger evidence for physics beyond $\lambda$cdm.
\newblock {\em The Astrophysical Journal}, 876(1):85, 2019.

\bibitem{Oyvind:2024axi}
Øyvind Grøn.
\newblock Can stiff matter solve the hubble tension?
\newblock {\em Axioms}, 13:526, 08 2024.

\end{thebibliography}

\end{document}